\documentclass[reprint, superscriptaddress, amsmath, amssymb, aps]{revtex4-2}

\usepackage{graphicx}
\usepackage{dcolumn}
\usepackage{bm}
\usepackage{xcolor}
\usepackage{mathtools}
\usepackage{physics}
\usepackage[normalem]{ulem}
\usepackage{lineno}

\usepackage{hyperref} 
\hypersetup{colorlinks=true, citecolor=blue}

\begin{document}

\title{Generalization of Wigner Time Delay to Sub-Unitary Scattering Systems}

\author{Lei Chen}
 \email{LChen95@umd.edu}
 \affiliation{Quantum Materials Center, Department of Physics, University of Maryland, College Park, MD 20742, USA}
 \affiliation{Department of Electrical and Computer Engineering, University of Maryland, College Park, MD 20742, USA}
\author{Steven M. Anlage}
 \email{anlage@umd.edu}
 \affiliation{Quantum Materials Center, Department of Physics, University of Maryland, College Park, MD 20742, USA}
 \affiliation{Department of Electrical and Computer Engineering, University of Maryland, College Park, MD 20742, USA}
\author{Yan V. Fyodorov}
\email{yan.fyodorov@kcl.ac.uk}
 \affiliation{Department of Mathematics, King’s College London, London WC26 2LS, United Kingdom}
 \affiliation{L. D. Landau Institute for Theoretical Physics, Semenova 1a, 142432 Chernogolovka, Russia}
 
\date{\today}
 
\begin{abstract}
We introduce a complex generalization of Wigner time delay $\tau$ for sub-unitary scattering systems. Theoretical expressions for complex time delay as a function of excitation energy, uniform and non-uniform loss, and coupling, are given. We find very good agreement between theory and experimental data taken on microwave graphs containing an electronically variable lumped-loss element. We find that time delay and the determinant of the scattering matrix share a common feature in that the resonant behavior in $\text{Re}[\tau]$ and $\text{Im}[\tau]$ serves as a reliable indicator of the condition for Coherent Perfect Absorption (CPA). This work opens a new window on time delay in lossy systems and provides a means to identify the poles and zeros of the scattering matrix from experimental data. The results also enable a new approach to achieving CPA at an arbitrary energy/frequency in complex scattering systems. 
\end{abstract}

\maketitle

\emph{Introduction.}  In this paper we consider the general problem of scattering from a complex system by means of excitations coupled through one or more scattering channels. The scattering matrix $S$ describes the transformation of a set of input excitations $\ket{\psi_\text{in}}$ on $M$ channels into the set of outputs $\ket{\psi_\text{out}}$ as $\ket{\psi_\text{out}}=S\ket{\psi_\text{in}}$.

A measure of how long the excitation resides in the interaction region is provided by the time delay, related to the energy derivative of the scattering phase(s) of the system. This quantity and its variation with energy and other parameters can provide useful insights into the properties of the scattering region and has attracted research attention since the seminal works by Wigner \cite{Wigner55} and Smith \cite{Smith60}. A review on theoretical aspects of time delays with emphasis to solid state applications can be found in \cite{Texier16}. Various aspects of time delay have recently been shown to be of direct experimental relevance for manipulating wave fronts in complex media \cite{Rotter11,Carpenter15,Horodynski20}.
Time delays are also long known to be directly related to the density of states of the open scattering system, see discussions in \cite{Texier16} and more recently in \cite{Kuipers14,Davy15}.

For the case of flux-conserving scattering in systems with no losses, the $S$-matrix is unitary and its eigenvalues are phases $e^{i\theta_a}, a = 1, 2, ..., M$. These phases are functions of the excitation energy $E$ and one can then define several different measures of time delay, see e.g. \cite{Fyodorov97,Texier16}, such as partial time delays associated with each channel $\tau_a=d\theta_a/dE$, the proper time delays which are the eigenvalues of the Wigner-Smith matrix $\hat{Q}=i\hbar \frac{dS^{\dagger}}{dE}S$, and the Wigner delay time which is the average of all the partial time delays ($\tau_\text{W} = \frac{1}{M} \sum_{a=1}^{M} \tau_a = \frac{1}{M} Tr[\hat{Q}]$). 

A rich class of systems in which properties of various time delays enjoyed thorough theoretical attention is scattering of short-wavelength waves from classically chaotic systems, e.g. billiards with ray-chaotic dynamics or particles on graphs, e.g. such as considered in \cite{BarraGaspard01}. Various examples of chaotic wave scattering (quantum or classical) have been observed in nuclei, atoms, molecules, ballistic two-dimensional electron gas billiards, and most extensively in microwave experiments \cite{Stock99,Richter01,Hul12,Kuhl13,Grad14,Dietz15}. In such systems time delays have been measured starting from the pioneering work \cite{Doron90}, followed over the last three decades by measurement of the statistical properties of time delay through random media \cite{Genack99,Genack03} and microwave billiards \cite{Schanze05}. Wigner time delay for an isolated resonance described by an $S$-matrix pole at complex energy $E_0-i\Gamma$ has a value of $Q=2\hbar/\Gamma$ on resonance, hence studies of the imaginary part of the $S$-matrix poles probe one aspect of time delay \cite{Kuhl08,Difa12,Bark13,Gros14,Liu14,DavyGenack18}. In the meantime, the Wigner-Smith operator (WSO) was utilized to identify minimally-dispersive principal modes in coupled multi-mode systems \cite{FanKahn05,Xiong16}. A similar idea was used to create particle-like scattering states as eigenstates of the WSO \cite{Rotter11,Gerar16,Bohm18}. A generalization of the WSO allowed maximal focus on, or maximal avoidance of, a specific target inside a multiple scattering medium \cite{Ambi17,Horodynski20}.
 
Time delays in wave-chaotic scattering are expected to be extremely sensitive to variations of excitation energy and scattering system parameters, and will display universal fluctuations when considering an ensemble of scattering systems with the same general symmetry. Universality of fluctuations allows them to be efficiently described using the theory of random matrices \cite{Lehmann95,Gopar96,Fyodorov97,Fyodorov97a,Brouwer99,SFS01,MezSimm13,Texier13,Novaes15,Cunden15}. Alternative theoretical treatments of time delay in chaotic scattering systems successfully adopted a semi-classical approach, see \cite{Kuipers14} and references therein.

Despite the fact that standard theory of wave-chaotic scattering deals with perfectly flux-preserving systems, in any actual realisation such systems are inevitably imperfect, hence absorbing, and theory needs to take this aspect into account \cite{FyoSavSomRev05}.
Interestingly, studying scattering characteristics in a system with weak uniform (i.e. spatially homogeneous) losses may even provide a possibility to extract time delays characterizing idealized system without losses. This idea has been experimentally realized already in \cite{Doron90} which treated the effect of sub-unitary scattering by means of the unitary deficit of the $S$-matrix. In this case consider the $Q$-matrix defined through the relation $S^{\dagger} S=1-(\gamma \Delta/2\pi) Q_{UD}$, where $\gamma$ is the dimensionless `absorption rate' and $\Delta$ is the mean spacing between modes of the closed system. In the limit of vanishing absorption rate $\gamma\to 0$ such $Q_{UD}$ can be shown to coincide with the Wigner-Smith time delay matrix for a lossless system, but formally one can extend this as a definition of $Q$ for any $\gamma>0$. Note that this version of time delay is always real and positive. Various statistical aspects of time delays in such and related settings were addressed theoretically in \cite{BeeBrouwer01,Fyodorov03,SavSomm03,Grabsch20}.
 
Experimental data is often taken on sub-unitary scattering systems and a straightforward use of the Wigner time delay definition yields a complex quantity. In addition, both the real and imaginary parts acquire both negative and positive values, and they show a systematic evolution with energy/frequency and other parameters of the scattering system. This clearly calls for a detailed theoretical understanding of this complex generalization of the Wigner time delay. It is necessary to stress that many possible definitions of time delays which are equivalent or directly related to each other in the case of a lossless flux-conserving systems can significantly differ in the presence of flux losses, either uniform or spatially localized. In the present paper we focus on a definition that can be directly linked to the fundamental characteristics of the scattering matrix - its poles and zeros in the complex energy plane, making it useful for fully characterizing an arbitrary scattering system. Note that $S$-matrix poles have been objects of long-standing theoretical \cite{SokZel89,Haake92,Fyod96,FyodKhor99,Somm99,fyod02,Poli09,Celardo11,Fyodorov2016} and experimental \cite{Kuhl08,Difa12,Bark13,Liu14} interest in chaotic wave scattering, whereas $S$-matrix zeroes started to attract research attention only recently \cite{Li2017,Fyodorov2017,BarOptica17,DavyGenack18,Fyo2019,Krasnok19,antilasing,OsmanFyo20,Chen2020,Imani20}.

\emph{Complex Wigner Time Delay.}  In our exposition we use the framework of the so-called ``Heidelberg Approach'' to wave-chaotic scattering reviewed from different perspectives in \cite{MRW2010,FSav11} and \cite{Schomerus2015}. 
Let $H$ be the $N \times N$ Hamiltonian which is used to model the closed system with ray-chaotic dynamics, $W$ denoting the $N \times M$ matrix of coupling elements between the $N$ modes of $H$ and the $M$ scattering channels, and by $A$ the $N \times L$ matrix of coupling elements between the modes of $H$ and the $L$ localized absorbers, modelled as $L$ absorbing channels. \footnote{This way of modelling the localized absorbers as additional scattering channels is close in spirit to the so-called dephasing lead model of decoherence introduced in: M. B{\"{u}}ttiker, Role of quantum coherence in series resistors, \href{https://doi.org/10.1103/PhysRevB.33.3020}{Phys. Rev. B \textbf{33}, 3020 (1986)} and further developed in P. W. Brouwer and C. W. J. Beenakker, Voltage-probe and imaginary-potential models for dephasing in a chaotic quantum dot, \href{https://doi.org/10.1103/PhysRevB.55.4695}{Phys. Rev. B \textbf{55}, 4695 (1997)}.} The total unitary $S$-matrix, of size $(M+L)\times (M+L)$ describing both the scattering and absorption on equal footing, has the following block form, see e.g. \cite{Fyodorov2017}:
\begin{align}
    \label{eq1}
    \mathcal{S}(E) &= 
    \begin{pmatrix}
    1_M - 2\pi iW^{\dagger}D^{-1}(E)W & - 2\pi iW^{\dagger}D^{-1}(E)A \\
    - 2\pi iA^{\dagger}D^{-1}(E)W & 1_L - 2\pi iA^{\dagger}D^{-1}(E)A
    \end{pmatrix}, 
\end{align}
where we defined $D(E)= E - H + i(\Gamma_W + \Gamma_A)$ with $\Gamma_W = \pi WW^{\dagger}$ and $\Gamma_A = \pi AA^{\dagger}$.

The upper left diagonal $M\times M$ block of $\mathcal{S}(E)$ is the experimentally-accessible sub-unitary scattering matrix and is denoted as $S(E)$. The presence of uniform-in-space absorption with strength $\gamma$ can be taken into account by evaluating the $S$-matrix entries at complex energy: $S(E+i\gamma) \coloneqq S_{\gamma}(E)$. The determinant of such a subunitary scattering matrix $S_{\gamma}(E)$ is then given by:
\begin{align}
    \label{eq3}
    \det S_{\gamma}(E) &\coloneqq \det S(E+i\gamma) \\
    \label{eq4}
    &= \frac{\det [E-H+i(\gamma + \Gamma_{A} - \Gamma_{W})]}
    {\det [E-H+i(\gamma + \Gamma_{A} + \Gamma_{W})]} \\
    \label{eq5}
    &= \prod_{n=1}^{N} \frac{E+i\gamma - z_n}{E+i\gamma - \mathcal{E}_n},
\end{align}

In the above expression we have used that the $S$-matrix zeros $z_n$ are complex eigenvalues of the non-self-adjoint/non-Hermitian matrix $H+i(\Gamma_W - \Gamma_A)$, whereas the poles $\mathcal{E}_n = E_n - i\Gamma_n$ with $\Gamma_n>0$ are complex eigenvalues of yet another non-Hermitian matrix $H-i(\Gamma_W + \Gamma_A)$, frequently called in the literature ``the effective non-Hermitian Hamiltonian'' \cite{SokZel89,Fyodorov97,Irotter09,FSav11,Schomerus2015,Fyodorov2016}. Note that when localized absorption is absent, i.e. $\Gamma_A=0$, the zeros $z_n$ and poles $\mathcal{E}_n$ are complex conjugates of each other, as a consequence of $S$-matrix unitarity for real $E$ and no uniform absorption $\gamma=0$.
Extending to locally absorbing systems the standard definition of the Wigner delay time as the energy derivative of the total phase shift we now deal with a complex quantity:

\begin{widetext}
\begin{align}
    \label{eq6}
    \tau(E;A,\gamma) &\coloneqq \frac{-i}{M} \frac{\partial}{\partial E} \log \det S_{\gamma}(E) \\
    \label{eq7}
    &= \text{Re}\ \tau(E;A,\gamma) + i\text{Im}\ \tau(E;A,\gamma), \\
    \label{eq8}
    \text{Re}\ \tau(E;A,\gamma) &= \frac{1}{M} \sum_{n=1}^{N} \left[ \frac{\text{Im}z_n - \gamma}{(E-\text{Re}z_n)^2 + (\text{Im}z_n - \gamma)^2} + \frac{\Gamma_n + \gamma}{(E-E_n)^2 + (\Gamma_n + \gamma)^2} \right], \\
    \label{eq9}
    \text{Im}\ \tau(E;A,\gamma) &= -\frac{1}{M} \sum_{n=1}^{N} \left[ \frac{E - \text{Re}z_n}{(E-\text{Re}z_n)^2 + (\text{Im}z_n - \gamma)^2} - \frac{E-E_n}{(E-E_n)^2 + (\Gamma_n + \gamma)^2} \right]
\end{align}
\end{widetext}

Equation (\ref{eq8}) for the real part is formed by two Lorentzians for each mode of the closed system, potentially with different signs. This is a striking difference from the case of the flux-preserving system in which the conventional Wigner time delay is expressed as a single Lorentzian for each resonance mode \cite{Lyub77}. Namely, the first Lorentzian is associated with the $n$th zero while the second is associated with the corresponding pole of the scattering matrix. The widths of the two Lorentzians are controlled by system scattering properties, and when $\text{Im}z_n \to \gamma \pm 0$ the first Lorentzian in Eq. \ref{eq8} acquires the divergent, delta-functional peak shape, of either positive or negative sign, centered at $E=\text{Re}z_n$. Note that the first term in Eq. \ref{eq9} changes its sign at the same energy value. These properties are indicative of the ``perfect resonance'' condition, with divergence in the real part of the Wigner time delay signalling the wave/particle being perpetually trapped in the scattering environment. In different words, the energy of the incident wave/particle is perfectly absorbed by the system due to the finite losses.

The pair of equations (\ref{eq8}, \ref{eq9}) forms the main basis for our consideration. In particular, we demonstrate in the Supp. Mat. Section I \cite{SuppMat} that in the regime of well-resolved resonances Eqs. (\ref{eq8}) and (\ref{eq9}) can be used for extracting the positions of both poles and zeros in the complex plane from experimental measurements, provided the rate of uniform absorption $\gamma$ is independently known. 
We would like to stress that in general the two Lorentzians in (\ref{eq8}) are centered at different energies because generically the pole position $E_n$ does not coincide with the real part of the complex zero $\text{Re}z_n$.

From a different angle it is worth noting that there is a close relation between the objects of our study and the phenomenon of the so called Coherent Perfect Absorption (CPA) which attracted considerable attention in recent years, both theoretically and experimentally \cite{CPA,Baran17,antilasing,Chen2020,Hougne20}. Namely, the above-discussed match between the uniform absorption strength and the imaginary part of scattering matrix zero $\gamma=\text{Im}z_n$ simultaneously ensures the determinant of the scattering matrix to vanish, see Eq. (\ref{eq5}). This is only possible when $\ket{\psi_\text{out}}=0$ despite the fact that $\ket{\psi_\text{in}}\neq 0$, 
which is a manifestation of CPA, see e.g. \cite{Li2017,Fyodorov2017}. 

\begin{figure*}[t]
\includegraphics[width=\textwidth]{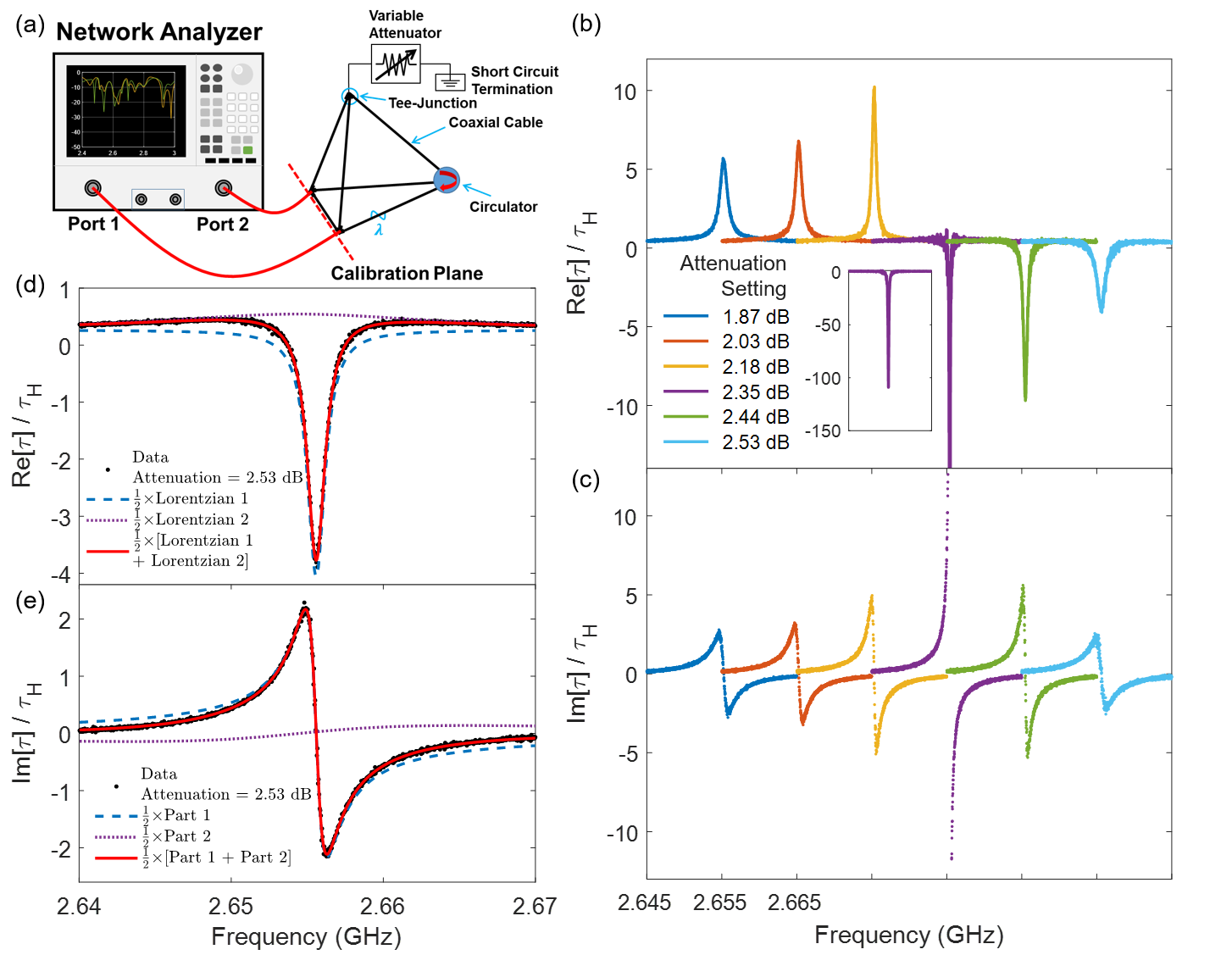}
\caption{(a) shows a schematic of the graph experimental setup. The lumped loss $\Gamma_A$ is varied by changing the applied voltage to the variable attenuator. (b) and (c) show experimental data of both real and imaginary parts of Wigner time delay $\text{Re}[\tau]$ and $\text{Im}[\tau]$ (normalized by the Heisenberg time $\tau_\text{H}$) as a function of frequency under different attenuation settings for a single isolated mode. For each attenuation setting, the data is plotted from 2.645 GHz to 2.665 GHz. For clarity, plots with higher attenuation setting are shifted 0.01 GHz from the previous one. Inset shows the entire range of $\text{Re}[\tau]$ for attenuation setting of 2.35 dB. (d) and (e) demonstrate the two-Lorentzian nature of the real and imaginary parts of the Wigner time delay as a function of frequency. The fitting parameters in these two plots are: $\text{Re}z_n=2.6556$ GHz, $E_n=2.6544$ GHz, $\text{Im}z_n-\gamma=-7.1065\times10^{-4}$ GHz, and $\Gamma_n+\gamma=0.0110$ GHz. The constants used in the $\text{Re}[\tau]$ and $\text{Im}[\tau]$ fits are $C_\text{R}=0.26$ and $C_\text{I}=-0.0018$ in units of $\tau_\text{H}$. Detailed discussion about the fitting constants and degree of isolation of the modes can be found in the Supp. Mat. section IV \cite{SuppMat}.}
\label{fig1}
\end{figure*}

\emph{Experiment.}  We focus on experiments involving microwave graphs \cite{Hul04,Lawn08,Hul12,Chen2020} for a number of reasons. First, they provide for complex scattering scenarios with well-isolated modes amenable to detailed analysis. We thus avoid the complications of interacting poles and related interference effects \cite{Pers98}. Graphs also allow for convenient parametric control such as variable lumped lossy elements, variable global loss, and breaking of time-reversal invariance. We utilize an irregular tetrahedral microwave graph formed by coaxial cables and Tee-junctions, having $M=2$ single-mode ports, and broken time-reversal invariance. A voltage-controlled variable attenuator is attached to one internal node of the graph (see Fig. \ref{fig1}(a)), providing for a variable lumped loss ($L=1$, the control variable $\Gamma_{A}$). The nodes involving connections of the graph to the network analyzer, and the graph to the lumped loss, are made up of a pair of Tee-junctions. The coaxial cables and tee-junctions have a roughly uniform and constant attenuation produced by dielectric loss and conductor loss, which is parameterized by the uniform loss parameter $\gamma$. The 2-port graph has a total electrical length of $L_e=3.89$ m, a mean mode spacing of $\Delta=c/2L_e=38.5$ MHz, and a Heisenberg time $\tau_\text{H}=2\pi/\Delta=$ 163 ns. The graph has equal coupling on both ports, characterized by a nominal value of $T_a = 0.9450$ at a frequency of 2.6556 GHz. \footnote{The coupling strength $T_a$ is determined by the value of the radiation $S$-matrix ($T_a=1-|S_{\text{rad}}|^2$). The radiation $S$ is measured when the graph is replaced by $50\ \Omega$ loads connected to the three output connectors of each node attached to the network analyzer test cables.}

\emph{Comparison of Theory and Experiments.}  Figure \ref{fig1} shows the evolution of complex time delay for a single isolated mode of the $M=2$ port tetrahedral microwave graph as $\Gamma_{A}$ is varied. The complex time delay is evaluated as in Eq. \ref{eq6} based on the experimental $S(f)$ data, where $f$ is the microwave frequency, a surrogate for energy $E$. Note that the (calibrated) measured S-parameter data is directly used for calculation of the complex time delay without any data pre-processing. The resulting real and imaginary parts of the time delay vary systematically with frequency, adopting both positive and negative values, depending on frequency and lumped loss in the graph. The full evolution animated over varying lumped loss is available in the Supplemental Material \cite{SuppMat}. These variations are well-described by the theory given above. 

Figure \ref{fig1}(d) and (e) clearly demonstrates that two Lorentzians are required to correctly describe the frequency dependence of the real part of the time delay. The two Lorentzians have different widths in general, given by the values of $\text{Im}z_n - \gamma$ and $\Gamma_n + \gamma$, and in this case the Lorentzians also have opposite sign. The frequency dependence of the imaginary part of the time delay also requires two terms, with the same parameters as for the real part, to be correctly described. The data in Fig. \ref{fig1}(b) also reveals that $\text{Re}[\tau]$ goes to very large positive values and suddenly changes sign to large negative values at a critical amount of local loss. For another attenuation setting of the same mode it was found that the maximum delay time was 337 times the Heisenberg time, showing that the signal resides in the scattering system for a substantial time.

\begin{figure}[ht]
\includegraphics[width=86mm]{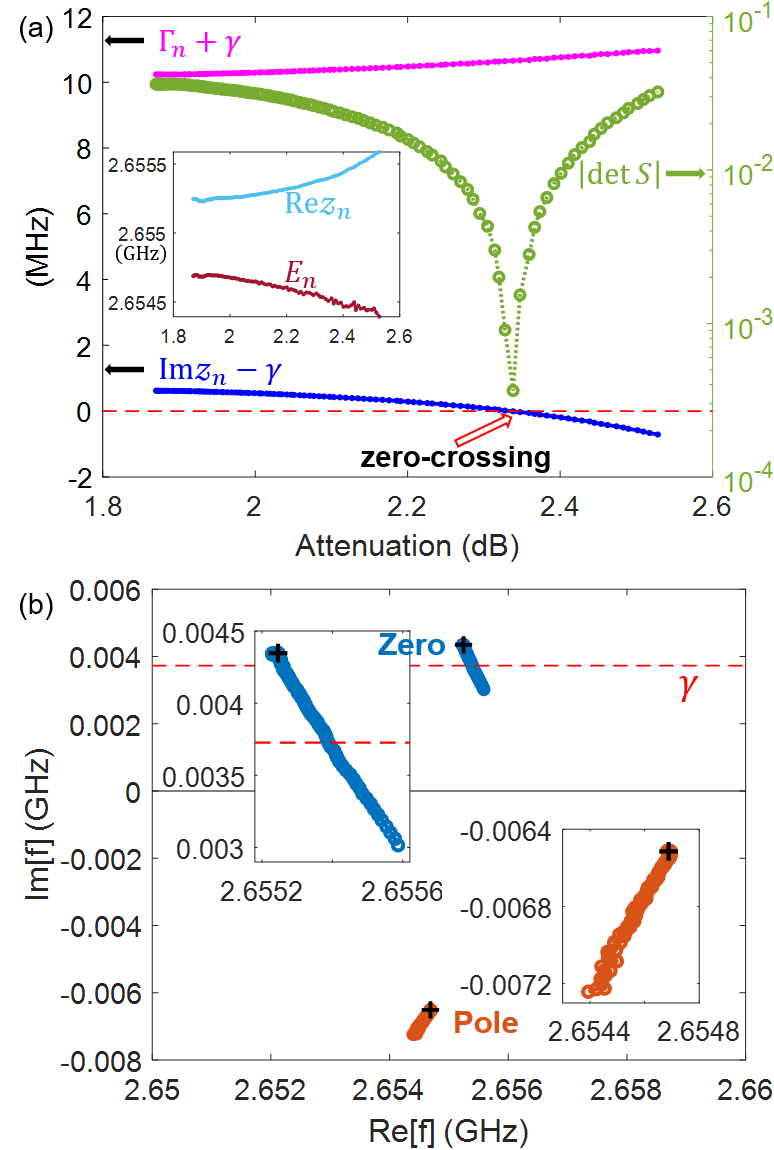}
\caption{(a) Fitted parameters $\text{Im}z_n-\gamma$ and $\Gamma_n+\gamma$ for the complex Wigner time delay from graph experimental data. Also shown is the evolution of $|\det (S)|$ at the specific frequency of interest, $f_\text{CPA}$, which reaches its minimum at the zero-crossing point. Inset shows the evolution of $\text{Re}z_n$ and $E_n=\text{Re}\mathcal{E}_n$ with attenuation. (b) Evolution of complex zero and pole of a single mode of the graph in the complex frequency plane as a function of $\Gamma_A$. The black crosses are the initial state of the zero and pole at the minimum attenuation setting. Insets show the details of the complex zero and pole migrations.}
\label{fig2}
\end{figure}

The measured complex time delay as a function of frequency can be fit to Eqs. (\ref{eq8}) and (\ref{eq9}) to extract the corresponding pole and zero location for the $S$-matrix. The method to perform this fit is described in the Supp. Mat. Section I \cite{SuppMat} The fitting parameters are $\text{Re}z_n$ and $\text{Im}z_n - \gamma$ for the zero, and $E_n$ and $\Gamma_n+\gamma$ for the pole. Note that the $\text{Re}[\tau(f)]$ and $\text{Im}[\tau(f)]$ data are fit simultaneously, and constant offsets $C_\text{R}$ and $C_\text{I}$ are added to each fit.

Figure \ref{fig2} summarizes the parameters required to fit the experimental complex time delay vs. frequency (shown in Fig. \ref{fig1}) as the localized loss due to the variable attenuator in the graph is increased. The significant feature here is the zero-crossing of $\text{Im}z_n - \gamma$ at frequency $f=f_\text{CPA}$, which corresponds to the point at which $\text{Re}[\tau(f)]$ changes sign. As shown in Fig. \ref{fig2}(a) this coincides with the point at which $|\det (S(f))|$ achieves its minimum value at the CPA frequency $f_\text{CPA}$. This demonstrates that one or more eigenvalues of the $S$-matrix go through a complex zero value precisely as the condition $\text{Im}z_n - \gamma=0$ and $f-\text{Re}z_n=0$ is satisfied. Associated with this condition $|\text{Re}[\tau(f_\text{CPA})]|$ diverges, with corresponding large positive and negative values of $\text{Im}[\tau(f)]$ occurring just below and just above $f=f_\text{CPA}$. Similar behavior of $\text{Re}[\tau(f)]$ was recently observed in a complex scattering system containing re-configurable metasurfaces, as the pixels were toggled \cite{Hougne20}.

Next we wish to estimate the value of uniform attenuation $\gamma$ for the microwave graph. Using the unitary deficit of the $S$-matrix in a setup in which the attenuator is removed \cite{Doron90}, we evaluate the uniform loss strength $\gamma$ to be $3.73 \times 10^{-3}$ GHz (see Supp. Mat. section III \cite{SuppMat}).

Figure \ref{fig2}(b) summarizes the locations of the $S$-matrix pole $\mathcal{E}_n$ and zero $z_n$ of the single isolated mode of the microwave graph in the complex frequency plane as the localized loss is varied. When the $S$-matrix zero crosses the $\text{Im}z_n = \gamma$ value, one has the traditional signature of CPA. Note from Fig. \ref{fig2} that the real parts of the zero and pole do not coincide and in fact move away from each other as localized loss is increased.



\emph{Discussion.}  It should be noted that the occurrence of a negative real part of the time delay is an inevitable consequence of sub-unitary scattering, and is also expected for particles interacting with attractive potentials \cite{Uzy17}. 

The imaginary part of time delay was in the past discussed in relation to changes in scattering unitary deficit with frequency \cite{Bohm18}. Another approach to defining complex time delay has been recently suggested to be based on essentially calculating the time delay of the signal which comes out of the system without being absorbed \cite{Hougne20}. It should be noted that this \emph{ad hoc} definition of time delay is not simply related to the poles and zeros of the $S$-matrix. Moreover, a closer inspection shows that such a definition of complex time delay tacitly assumes that the real parts of the pole and zero are identical. According to our theory such an assumption is incompatible with a proper treatment of localized loss.

We emphasize that the correct knowledge of the locations of the poles and zeros is essential for reconstructing the scattering matrix over the entire complex energy plane through Weierstrass factorization \cite{Grig13}.
Through graph simulations presented in Sup. Mat. Section VII \cite{SuppMat} we demonstrate that the complex time delay theory presented here also works for time-reversal invariant systems, and for systems with variable uniform absorption strength $\gamma$. Our results therefore establish a systematic procedure to find the $S$-matrix zeros and poles of isolated modes of a complex scattering system with an arbitrary number of coupling channels, symmetry class, and arbitrary degrees of both global and localized loss.

Recent work has demonstrated CPA in disordered and complex scattering systems \cite{antilasing,Chen2020}. It has been discovered that one can systematically perturb such systems to induce CPA at an arbitrary frequency \cite{Frazier20,Hougne20}, and this enables a remarkably sensitive detector paradigm \cite{Hougne20}. These ideas can also be applied to optical scattering systems where measurement of the transmission matrix is possible \cite{Pop10}. Here we have uncovered a general formalism in which to understand how CPA can be created in an arbitrary scattering system. In particular this work shows that both the global loss ($\gamma$), localized loss centers, or changes to the spectrum can be independently tuned to achieve the CPA condition. 

Future work includes treating the case of overlapping modes, and the development of theoretical predictions for the statistical properties of both the real and imaginary parts of the complex time delay in chaotic and multiple scattering sub-unitary systems.

\emph{Conclusions.}  We have introduced a complex generalization of Wigner time delay which holds for arbitrary uniform/global and localized loss, and directly relates to poles and zeros of the scattering matrix in the complex energy/frequency plane. Based on that we developed theoretical expressions for complex time delay as a function of energy, and found very good agreement with experimental data on a sub-unitary complex scattering system. Time delay and $\det (S)$ share a common feature that CPA and the divergence of $\text{Re}[\tau]$ and $\text{Im}[\tau]$ coincide. This work opens a new window on time delay in lossy systems, enabling extraction of complex zeros and poles of the $S$-matrix from data. 

We acknowledge Jen-Hao Yeh for early experimental work on complex time delay. This work was supported by AFOSR COE Grant No. FA9550-15-1-0171, NSF DMR2004386, and ONR Grant No. N000141912481.

\bibliography{WTD.bib}

\end{document}


\title{SUPPLEMENTARY MATERIAL for \\
Generalization of Wigner Time Delay to Sub-Unitary Scattering Systems}

\author{Lei Chen}
 \affiliation{Quantum Materials Center, Department of Physics, University of Maryland, College Park, MD 20742, USA}
 \affiliation{Department of Electrical and Computer Engineering, University of Maryland, College Park, MD 20742, USA}
\author{Steven M. Anlage}
 \affiliation{Quantum Materials Center, Department of Physics, University of Maryland, College Park, MD 20742, USA}
 \affiliation{Department of Electrical and Computer Engineering, University of Maryland, College Park, MD 20742, USA}
\author{Yan V. Fyodorov}
 \affiliation{Department of Mathematics, King’s College London, London WC26 2LS, United Kingdom}
 \affiliation{L. D. Landau Institute for Theoretical Physics, Semenova 1a, 142432 Chernogolovka, Russia}
 
\date{\today}

\maketitle

\renewcommand{\thefigure}{S\arabic{figure}}
\renewcommand{\theequation}{S\arabic{equation}}

We provide some additional details for some of the calculations described in the text of the Letter.

\section{Extracting poles and zeros from experimental data}

Consider a pair of single terms in the sum over $n$ in Eqs. (7--8)

\begin{equation} \label{App1}
\text{Re}\,\tau_n(f) = \left[\frac{\text{Im} z_{n}-\gamma}{(f-\text{Re} z_{n})^{2}+(\text{Im} z_{n}-\gamma)^{2}}+\frac{\Gamma_{n}+\gamma}{(f-E_{n})^{2}+(\Gamma_{n}+\gamma)^{2}}\right],
\end{equation}
\begin{equation}\label{App2}
    -\text{Im}\,\tau_n(f) = \left[\frac{f-\text{Re} z_{n}}{(f-\text{Re} z_{n})^{2}+(\text{Im} z_{n}-\gamma)^{2}}-\frac{f-E_{n}}{(f-E_{n})^{2}+(\Gamma_{n}+\gamma)^{2}}\right]
\end{equation}

where we `relabeled' the energy parameter $E$ as `frequency' $f$.

Extracting the parameters $\text{Re} z_{n}, \text{Im} z_{n}, E_n, \Gamma_n$ and the uniform absorption strength $\gamma$ from the experimentally measured curves $\text{Im}\,\tau_n(f)$ and $\text{Re}\,\tau_n(f)$ is possible by the following procedure.

\begin{enumerate}
\item Let us {\bf define} the ``resonance frequency'' value $f=f^*$ by the condition
\begin{equation}\label{resofreq}
 \text{Im}\,\tau_n(f^*)=0
\end{equation}
\item Measure from the two above curves the following $3$ parameters:
\begin{equation}\label{curvepar}
s=-\frac{d}{df}\text{Im}\,\tau_n(f)|_{f=f^*}, \quad  m=\text{Re}\,\tau_n(f^*), \quad  \delta=-\frac{1}{2}\frac{d}{df}\ln{\left[\text{Re}\,\tau_n(f)\right]}|_{f=f^*}
\end{equation}
and use them to build combinations:
\begin{equation}\label{curveparalpha}
 \alpha_{-}=\frac{1}{2}\left(-\frac{s}{m}+m\right), \quad  \alpha_{+}=\frac{1}{2}\left(\frac{s}{m}+m\right)
\end{equation}
The following relations then can be derived for the positions of resonance poles and zeros. 

For the real parts:
\begin{equation}\label{RealParts}
E_n=f^*-\frac{\delta}{\delta^2+\alpha_{-}^2}, \quad \text{Re}\,z_n=f^*-\frac{\delta}{\delta^2+\alpha_{+}^2}
\end{equation}
For the imaginary parts:
\begin{equation}\label{ImParts}
\Gamma_n=-\gamma+\frac{\alpha_{-}}{\delta^2+\alpha_{-}^2}, \quad \text{Im}\,z_n=\gamma+\frac{\alpha_{+}}{\delta^2+\alpha_{+}^2}
\end{equation}

\centerline{ Derivation of relations (\ref{RealParts})--(\ref{ImParts})}

Define for brevity $x_n:=f^{*}-\text{Re}{z_n}$, $r_n:=\text{Im}{z_n}-\gamma$, $\epsilon_n:=f^{*}-E_n$ and $p_n:=\Gamma_n+\gamma$.

Then condition (\ref{resofreq}) implies
\begin{equation}\label{A}
\frac{x_n}{x_n^2+r_n^2}=\frac{\epsilon_n}{\epsilon_n^2+p_n^2}
\end{equation}
Also we have by the definition
\begin{equation}\label{Reii}
m=\text{Re}\,\tau_n(f)|_{f=f^*}=\frac{r_n}{x_n^2+r_n^2}+\frac{p_n}{\epsilon_n^2+p_n^2}
\end{equation}

On the other hand differentiating (\ref{App2}) gives with this notation:
\begin{equation} \label{s1}
s=-\frac{d}{df}\text{Im}\,\tau_n(f)|_{f=f^*}=\frac{r_n^2-x^2_n}{(x_n^2+r_n^2)^2}-\frac{p_n^2-\epsilon^2_n}{(p_n^2+\epsilon_n^2)^2}
\end{equation}
Now using the condition (\ref{A}) the above equation simplifies to
\begin{equation}\label{s2}
s=\frac{r_n^2}{(x_n^2+r_n^2)^2}-\frac{p_n^2}{(p_n^2+\epsilon_n^2)^2}
=\left[\frac{r_n}{(x_n^2+r_n^2)}-\frac{p_n}{(p_n^2+\epsilon_n^2)}\right]
\left[\frac{r_n}{(x_n^2+r_n^2)}+\frac{p_n}{(p_n^2+\epsilon_n^2)}\right]
\end{equation}
which after using (\ref{Reii}) gives
\begin{equation}\label{fracImRe}
s/m=\left[\frac{r_n}{(x_n^2+r_n^2)}-\frac{p_n}{(p_n^2+\epsilon_n^2)}\right]
\end{equation}
Now the pair (\ref{Reii})--(\ref{fracImRe}) implies
\begin{equation}\label{B}
\alpha_+:=\frac{1}{2}\left(\frac{s}{m}+m\right)=\frac{r_n}{x_n^2+r_n^2}, \quad \alpha_{-}:=\frac{1}{2}\left(-\frac{s}{m}+m\right)=\frac{p_n}{p_n^2+\epsilon_n^2}
\end{equation}
Finally, let us consider
\begin{equation}\label{curvepar2}
-\frac{1}{2}\frac{d}{df}\text{Re}\,\tau_n(f)|_{f=f^*}=\frac{r_nx_n}{(x_n^2+r_n^2)^2}+\frac{p_n\epsilon_n}{(p_n^2+\epsilon_n^2)^2}
\end{equation}
which after first using (\ref{A}) and then (\ref{Reii}) can be further rewritten as
\[
=\frac{\epsilon_n}{p_n^2+\epsilon_n^2}\left[\frac{r_n}{x_n^2+r_n^2}+\frac{p_n}{p_n^2+\epsilon_n^2}\right]=\frac{\epsilon_n}{p_n^2+\epsilon_n^2}
\text{Re}\,\tau_n(f)|_{f=f^*}
\]
implying finally
\begin{equation}\label{C}
\delta:=-\frac{1}{2}\frac{d}{df}\ln{\left[\text{Re}\,\tau_n(f)\right]}|_{f=f^*}=\frac{\epsilon_n}{p_n^2+\epsilon_n^2}=\frac{r_n}{x_n^2+r_n^2}
\end{equation}
The pair of equations (\ref{B}) and (\ref{C}) can be easily solved and gives (\ref{RealParts})--(\ref{ImParts}).

So the only parameter which remains to be found from independent measurement is the uniform absorption $\gamma$. \\[1ex]

{\bf Note:} Note that generically both $s=-\frac{d}{df}\text{Im}\,\tau_n(f)|_{f=f^*}\ne 0$ 
and $m=\text{Re}\,\tau_n(f)|_{f=f^*}\ne 0$ so $|\alpha_{+}|\ne |\alpha_{-}|$. We then see from Eq. (\ref{RealParts}) that to have $E_n =\text{Re}\,z_n$ is only possible if $\delta=0$, which in turn is only possible if
 $\frac{d}{df}\left[\text{Re}\,\tau_n(f)\right]|_{f=f^*}=0$. The latter condition would mean that the maximum or minimum on the curve $\text{Re}\,\tau_n(f)$ happens exactly at the same frequency $f^*$ where the imaginary part vanishes. Generically this never happens due to the two-Lorentzian nature of $\text{Re}\,\tau_n(f)$. 

\end{enumerate}

\section{Sign Convention for the Phase Evolution of the $S$-Matrix Elements}
It should be noted that there are two widely-used conventions for the evolution of the phase of the complex $S$-matrix elements with increasing frequency. Microwave network analyzers utilize a convention in which the phase of the scattering matrix elements \emph{decreases} with increasing frequency. Here we adopt the convention used in the theoretical literature that the phase of $S$-matrix elements \emph{increases} with increasing frequency. 

\section{Evaluation of System Uniform Loss Strength $\gamma$}
\begin{figure}[ht]
\includegraphics[width=\textwidth]{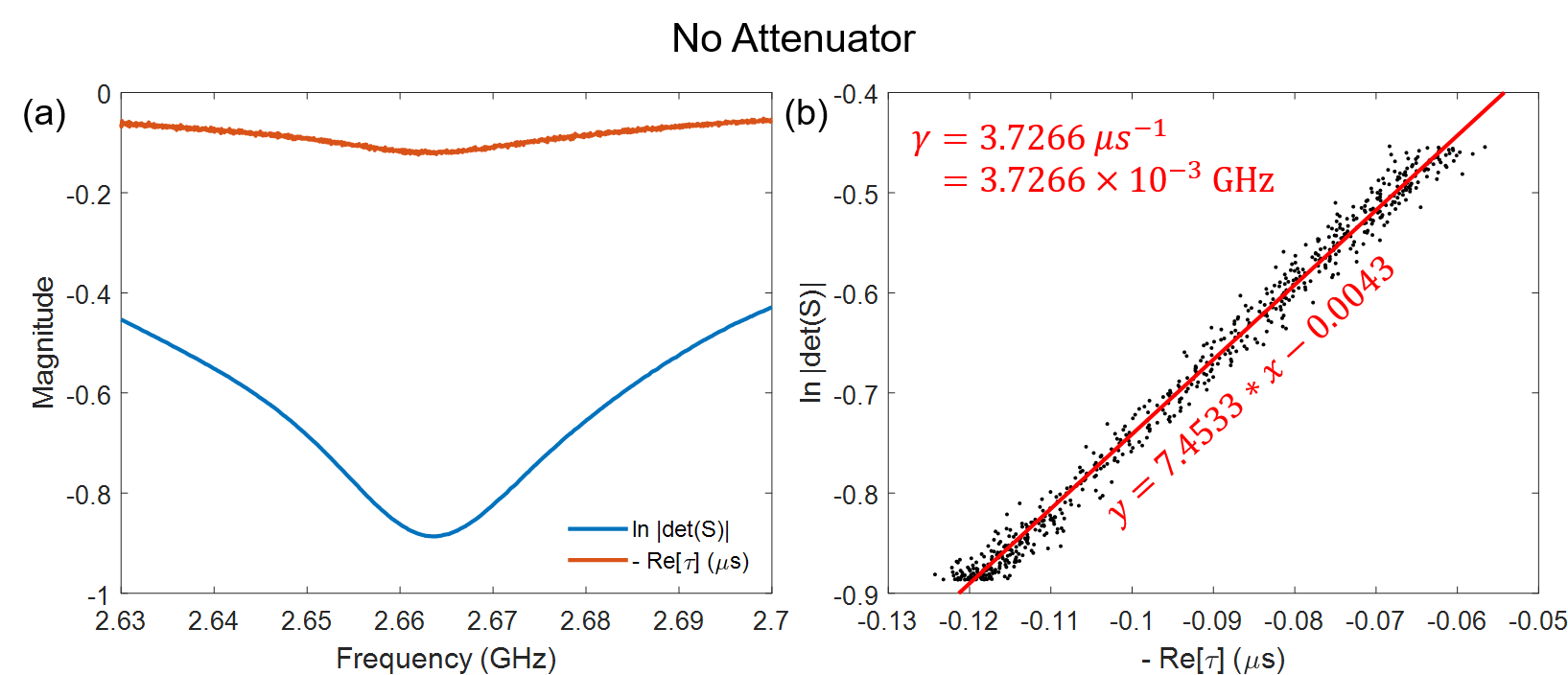}
\caption{(a) Measurement of $\ln{|\det S|}$ and $-\text{Re}[\tau]$ as a function of frequency for the resonance of interest in a no-attenuator tetrahedral graph. (b) Plot of $\ln{|\det S|}$ vs. $-\text{Re}[\tau]$ (with frequency as a parameter) and a linear fit between these two quantities to evaluate the uniform loss strength $\gamma$.}
\label{figS1}
\end{figure}

The value of $\gamma$ is estimated in a situation in which the uniform attenuation of the coaxial cables dominates the overall loss of the system. These losses arise from metallic and dielectric loss in the cables and are quite homogeneously distributed in the system. Here we estimate the value of uniform attenuation $\gamma$ for the microwave graph using the unitary deficit of the $S$-matrix \cite{Doron90}. We measured the graph with the variable attenuator removed (see Fig. 1(a) in the main text), and measured the $2\times2$ $S$-matrix under this no-attenuator condition. The unitary deficit of the $S$-matrix is expected to be $\ln{|\det S|} \approx -M\gamma\text{Re}[\tau]$ in the limit of low loss \cite{Doron90}, where $M$ is the number of channels ($M=2$ in this paper). Fig. \ref{figS1} shows the $\det S$ and $\text{Re}[\tau]$ versus frequency experimental data for the resonance of interest when no attenuator is present. The linear fitting function in Fig. \ref{figS1}(b) determines the uniform loss strength to be $\gamma = 3.73 \times 10^{-3}$ GHz.

\section{Effects of Neighboring Resonances}
\begin{figure}[ht]
\includegraphics[width=0.6\textwidth]{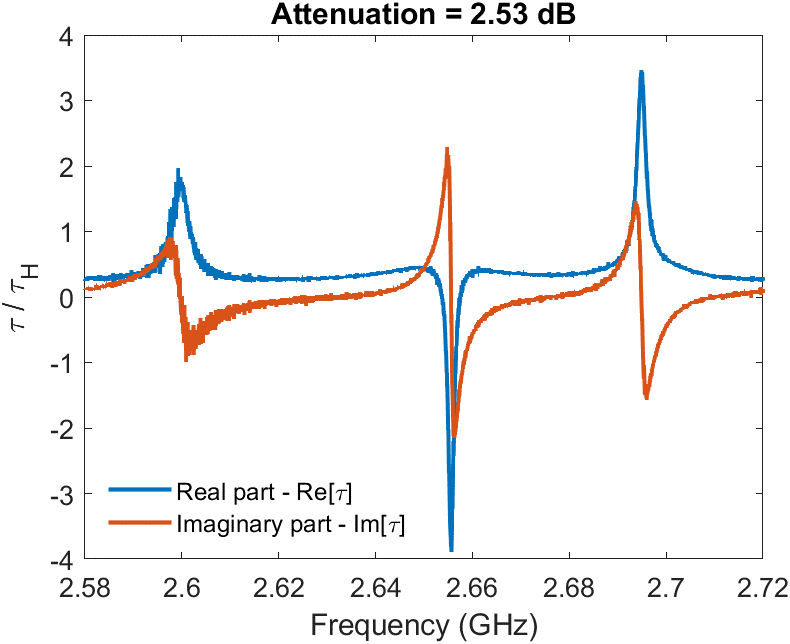}
\caption{Complex time delay experimental data for the neighboring resonances at an attenuation setting of 2.53 dB. The resonance mode in the middle is the one being analyzed in the main text.}
\label{figS2}
\end{figure}

In this paper we use an isolated mode for data fitting and analysis. A good definition of `isolated resonance' is that the separation is much larger than the width of the resonance. Fig. \ref{figS2} shows the separation between the mode fit in the main text and its neighboring resonances, where the nearest-neighbor distance for the middle resonance is $\Delta f=0.0393$ GHz (with the resonance on the right). Under this attenuation setting, the resonance widths of the two Lorentzians from the zero and pole for the middle resonance are $|\text{Im}z_n-\gamma|=7.1065\times10^{-4}$ GHz and $\Gamma_n+\gamma=0.0110$ GHz, which give rise to dimensionless separation estimates of $\frac{\Delta f}{|\text{Im}z_n-\gamma|}=55.3$ and $\frac{\Delta f}{\Gamma_n+\gamma}=3.6$, respectively.

\begin{figure}[ht]
\includegraphics[width=\textwidth]{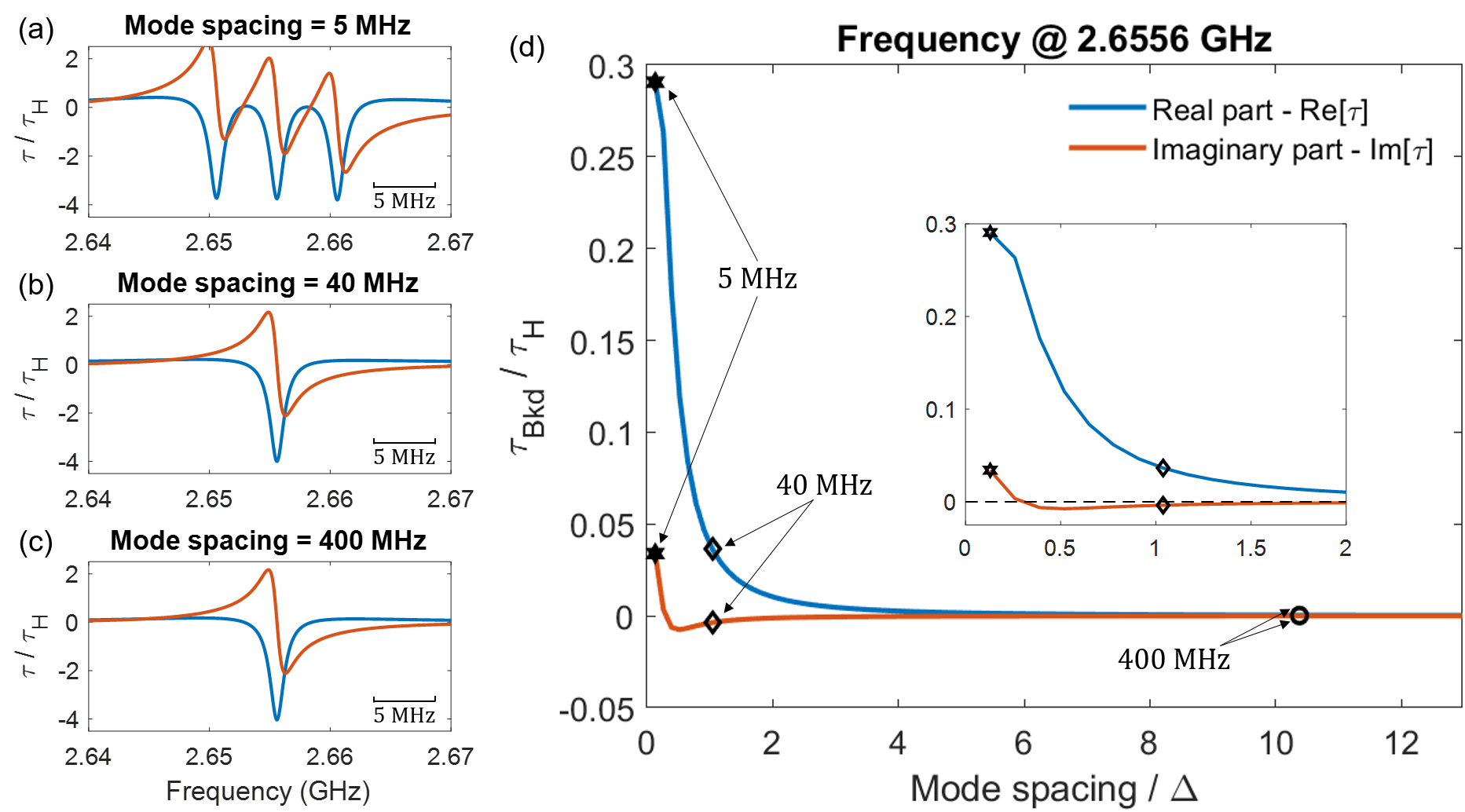}
\caption{Simulations based on Eqs. (7) and (8) of the main text of the effect of neighboring resonances on the background values $\tau_\text{Bkd}$ of $\text{Re}[\tau]$ and $\text{Im}[\tau]$ at the location of the center resonance. (a)--(c) show the complex time delay for three constructed neighboring resonances with variable mode spacing. A scale bar of 5 MHz is added for reference. Here $\tau_\text{H}=$163 ns is the Heisenberg time from the experiment discussed in the main text. The parameters used for the center resonance are: $\text{Re}z_n=2.6556$ GHz, $E_n=2.6544$ GHz, $\text{Im}z_n-\gamma=-7.1065\times10^{-4}$ GHz, and $\Gamma_n+\gamma=0.0110$ GHz. (d) shows the background contributions $\tau_\text{Bkd}$ of both the real and imaginary parts of complex time delay from the neighboring two resonances at the center frequency of 2.6556 GHz. Here $\Delta=38.5$ MHz is the mean mode spacing of the experimental graph in the main text, and is simply used as a characteristic frequency scale for normalization. The corresponding data points from (a) -- (c) are labelled in the plot. The background contributions decrease dramatically as the mode spacing increases, and the background contribution to $\text{Im}[\tau]$ is much smaller compared with the contribution to $\text{Re}[\tau]$. Inset in (d) shows a zoom-in view of the two contributions for small mode spacing.}
\label{figS3}
\end{figure}

In situations where the isolation between neighboring resonances is not large, the nearby resonances may have some contributions to the complex time delay at the resonance of interest. Here we evaluate this effect by using Eqs. (7) and (8) of the main text for three hypothetical modes. We utilize three similar pairs of zero and pole with adjustable mode spacing between them (see Fig. \ref{figS3}). The middle resonance has the same zero and pole from the experimental data, and the neighboring two resonances have the same information except for the variable frequency separation between them. Fig. \ref{figS3} clearly demonstrates that the neighboring resonances can have strong background contributions to $\text{Re}[\tau]$, which may lead to a non-negligible constant $C_\text{R}$ in the fitting process (see Fig. 1(d) and (e) in the main text and Fig. \ref{figS4}(d) and (e) in the Supp. Mat. section VII). The contribution to $\text{Im}[\tau]$, on the other hand, is quite small because $\text{Im}[\tau](f)$ changes sign through the resonance, hence for a low density of modes the `tails' of the $\text{Im}[\tau]$ contributions cancel out to good approximation. Fig. \ref{figS3}(d) shows the background contribution $\tau_\text{Bkd}$ to $\text{Re}[\tau]$ and $\text{Im}[\tau]$ at the resonance of interest as a function of the mode separation. Clearly $\text{Re}[\tau]$ is more sensitive to the presence of nearby modes and requires larger fit values for $C_\text{R}$, as compared to $C_\text{I}$, consistent with the results in Fig. 1 of the main text.

\section{Further Details about CPA and Complex Time Delay}
We note that at CPA both the peak in $|\text{Re}\ \tau|$ and the point at which $\text{Im}\ \tau$ changes its sign coincide in energy, but away from CPA these features may occur at different energies. Note that the real and imaginary parts share the same forms in the denominator of the Lorentzians. This leads to a synchronous evolution of their shapes with energy and scattering characteristics. This property is clearly demonstrated in both the data shown in the main text and in the simulations below in Section VII.

We also note that CPA can be achieved by tuning either localized loss $\Gamma_A$, or uniform loss $\gamma$, as demonstrated in Fig. 2 of the main text and Fig. \ref{figS5} of Supp. Mat. Section VII, respectively.

\section{Connections to Earlier Work on Negative Real Time Delay and Imaginary Time Delay}
In the past, the concept of a complex transmission delay was developed in the context of principal modes in multi-mode waveguides \cite{FanKahn05}, and a similar quantity was later used in the experimental realization of particle-like scattering states \cite{Bohm18}. Complex dwell time was defined for a multiple scattering medium with lossy resonant absorbers \cite{Durand19}.

An early study of Gaussian pulse propagation through an anomalously dispersive medium predicted negative delay \cite{Garr70}, and later measurements confirmed the theory \cite{Chu82}. The observed negative delay of the pulse was attributed to the fact that the leading edge of the pulse is attenuated less than the later parts, and the Gaussian shape is approximately preserved, under appropriate circumstances. Negative real parts of reflection delay time have been measured experimentally in a one-dimensional Levy-flight system but were dismissed as an artefact due to prompt reflections \cite{Razo20}.

 Our findings show that in general the real part of the delay as a function of frequency is not symmetric about the resonance as a consequence of the differences in the real parts of $z_n$ and $\mathcal{E}_n$, and that it has a distinctive two-Lorentzian character that had not been appreciated until now.
 
\section{Simulations of Graphs and Evaluation of Complex Time Delay}
We have simulated the microwave graphs using CST Microwave Studio utilizing an idealized simulation model. In this model, an irregular tetrahedral graph similar to that used in the experiment is considered. The graph nodes are represented by Tee-junctions which are set to be ideal (point-like with an ideal scattering matrix), and the only source of loss in the setup comes from uniform dielectric loss of the coaxial cables which can be conveniently varied by changing the dielectric loss parameter $\text{tan}\delta$. The 2-port graph simulation model has a total electrical length of 8.69 m, Heisenberg time $\tau_\text{H}=$ 364 ns, equal coupling on both ports, characterized by a nominal value of $T_a = 0.75$, and no lumped loss (i.e. $\Gamma_{A}=0$).  Note that in contrast to the experiment discussed in the main text, here we consider a time-reversal invariant microwave graph.  This is done, in part, to demonstrate that the complex time delay theory applies to all classes of complex scattering systems. 

\begin{figure*}[t]
\includegraphics[width=\textwidth]{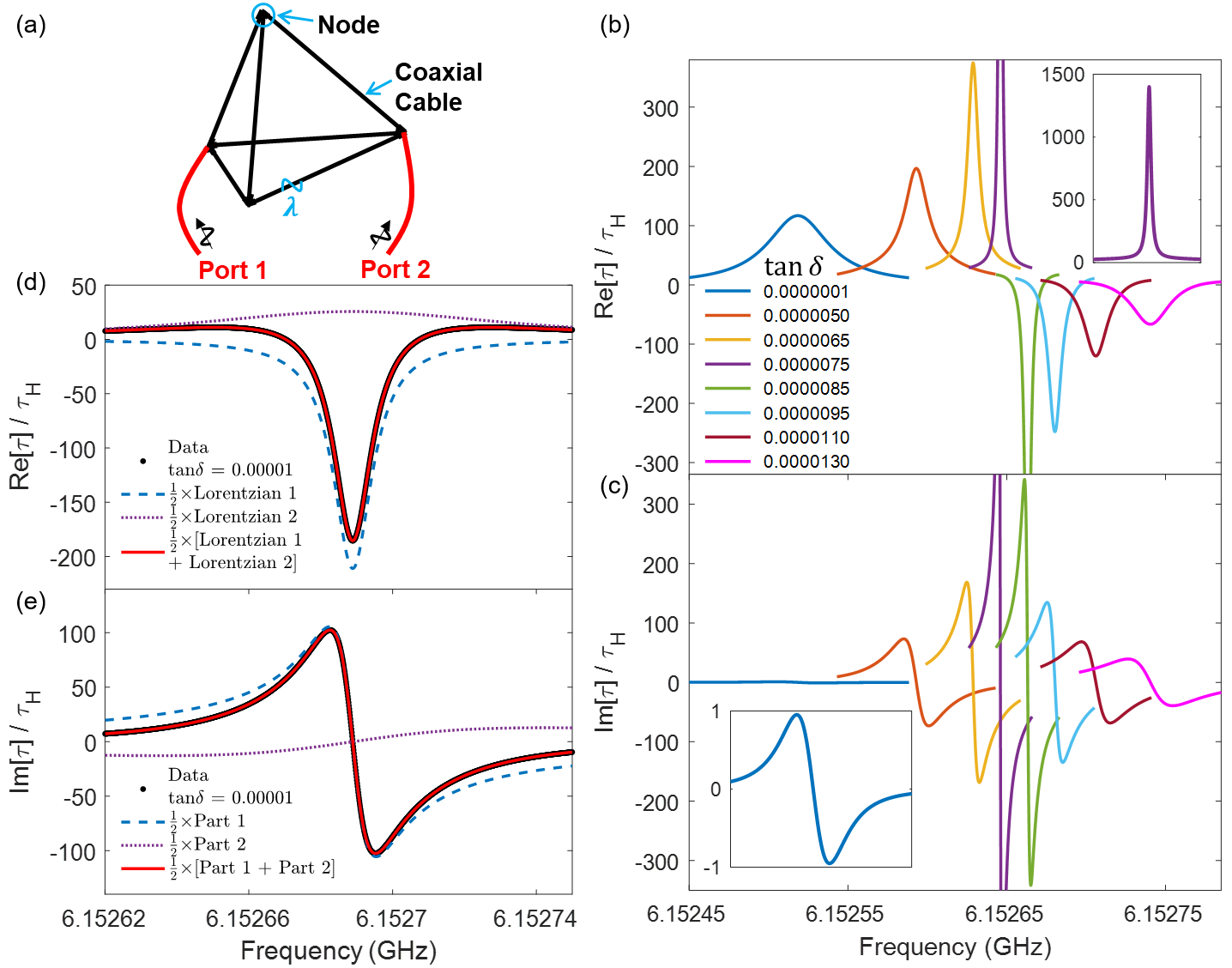}
\caption{Complex Wigner time delay from graph simulation with varying uniform loss $\gamma$. (a) shows a schematic of the graph simulation setup. The uniform loss $\gamma$ is varied by changing the dielectric loss parameter $\text{tan}\delta$ of the coaxial cables. (b) and (c) show simulation data of both real and imaginary parts of Wigner time delay $\text{Re}[\tau]$ and $\text{Im}[\tau]$ (normalized by the Heisenberg time $\tau_\text{H}$), as a function of frequency under different uniform loss settings for a single isolated mode near 6.1526 GHz. Inset in (b) shows the entire range of $\text{Re}[\tau]$ for $\text{tan}\delta=7.5\times10^{-6}$, and inset in (c) shows the zoom-in view of $\text{Im}[\tau]$ for $\text{tan}\delta=10^{-7}$. (d) and (e) demonstrate the two-Lorentzian nature of the real and imaginary parts of the Wigner time delay (normalized by the Heisenberg time $\tau_\text{H}$) as a function of frequency. The fitting parameters in these two plots are: $\text{Re}z_n=6.1527$ GHz, $E_n=6.1527$ GHz, $\text{Im}z_n-\gamma=-6.5057\times10^{-6}$ GHz, and $\Gamma_n+\gamma=5.3547\times10^{-5}$ GHz. The constants used in the $\text{Re}[\tau]$ and $\text{Im}[\tau]$ fits are $C_\text{R}=0.097$ and $C_\text{I}=0.00016$ in units of $\tau_\text{H}$.}
\label{figS4}
\end{figure*}

\begin{figure}[ht]
\includegraphics[width=0.6\textwidth]{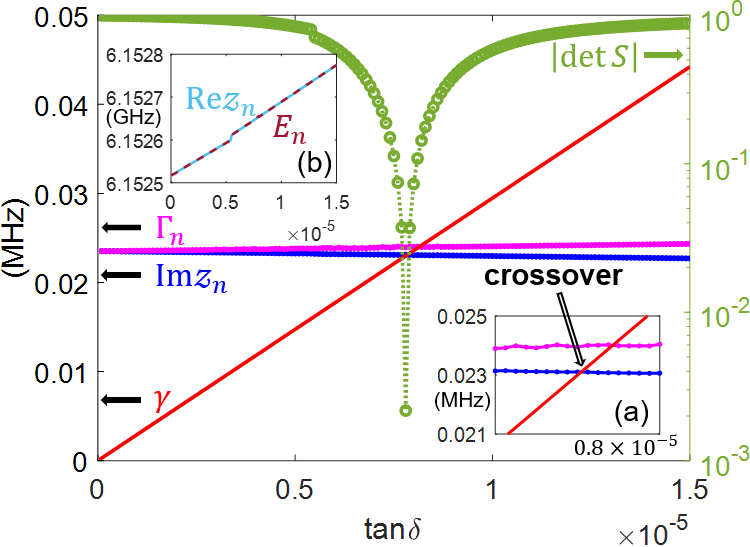}
\caption{Evolution of scattering matrix zero and pole for a single mode in the graph simulation with varying uniform loss $\gamma$. Plot shows the fitted parameters $\text{Im}z_n$ and $\Gamma_n$ for the complex Wigner time delay from graph simulation data. Also shown is the evolution of $|\det (S)|$ at the specific frequency of interest, $f_\text{CPA}$, which reaches its minimum at the crossover point where $\text{Im}z_n-\gamma=0$. Inset (a) shows the zoom-in details of the crossover when $\text{Im}z_n$ matches with $\gamma$, and inset (b) shows the evolution of $\text{Re}z_n$ and $E_n=\text{Re}\mathcal{E}_n$ with varying $\text{tan}\delta$. In this case, the real part of zero and pole are equal.}
\label{figS5}
\end{figure}

The new insight created by the simulation comes from the ability to systematically vary the uniform loss $\gamma$ while keeping all other parameters fixed. In this way we can show how CPA can be accomplished by simply changing the uniform loss. Results are shown in Figs. \ref{figS4} and \ref{figS5}, which are similar to the experimental results Figs. 1 and 2 in the main text. The CPA condition is achieved for a single isolated mode at $f_\text{CPA}=$ 6.15265 GHz when the uniform loss tangent value is tuned through a value of $\text{tan}\delta=7.8 \times 10^{-6}$.

\section{Animations of Time Delay Evolution with Variation of System Parameters}
Animation of experimental $\text{Re}[\tau(f)]$ and $\text{Im}[\tau(f)]$ as lumped loss is varied through the CPA condition.

Animation of simulation $\text{Re}[\tau(f)]$ and $\text{Im}[\tau(f)]$ as uniform loss $\gamma$ is varied through the CPA condition.

\bibliography{WTD.bib}